# $\boldsymbol{Smooth}$ $\%\boldsymbol{MinMax}$: A Differentiable Relaxation for Codon Harmonization

*Yoonho Jeong,*[†] *Hyunwoo Choi,*[†] *Ryan Fernandez Medina Hariri,*[‡] *Eok Kyun Lee,*[†] *Seung Seo Lee,*[‡] *and Insung S. Choi*[†,*]

[†]Department of Chemistry, KAIST, Daejeon 34141, Korea.
[‡]School of Chemistry and Chemical Engineering, Highfield Campus, University of Southampton, Southampton SO17 1BJ, United Kingdom.

**Abstract:** Codon harmonization aims to adapt the coding sequences for heterologous expression while preserving the native-like patterns of frequent and rare codons that may influence local translation dynamics and co-translational protein folding. However, widely used harmonization metrics, such as $\%MinMax$, are defined on discrete codon sequences and are, therefore, not readily compatible with the gradient-based neural codon design. Here, we introduce $\boldsymbol{Smooth}$ $\%\boldsymbol{MinMax}$, denoted as $\%\boldsymbol{MinMax}_{[s]}$, a differentiable relaxation of the conventional hard $\%MinMax$ metric, denoted as $\%MinMax_{[h]}$. $\%MinMax_{[s]}$ replaces the discrete codon-usage values with probability-weighted synonymous-codon usage values and does the hard $\%Max/\%Min$ branch with a sigmoid-gated interpolation. This formulation preserves the signed interpretation of $\%MinMax_{[h]}$, while enabling optimization with respect to the synonymous-codon probabilities and learnable parameters. In human-to-*Escherichia coli* codon harmonization experiments, $\%MinMax_{[s]}$ closely approximates $\%MinMax_{[h]}$ and supports the gradient-based profile matching in synonymous-codon probability space. Moreover, profile-guided beam search converted the learned codon probabilities into discrete codon sequences while maintaining close agreement with the target profiles. These results suggest $\%MinMax_{[s]}$ as a practical bridge between profile-based codon harmonization and neural synonymous-sequence design.

## 1. Introduction

The choice among synonymous codons is an important design consideration in the heterologous gene expression, in which a gene from one organism is expressed in a different host organism. Although synonymous codons encode the same amino acid, they are not used equally across species. Instead, each organism exhibits the characteristic patterns of "codon-usage bias", reflecting differences in tRNA abundance, genome composition, evolutionary history, the properties of the cellular translation machinery, etc.[1-6] As a result, a coding sequence, which is efficiently translated in its native organism, may not be expressed with the same efficiency in a heterologous host.

The classical approach of codon optimization typically replaces the codons that are rare in the expression host with the synonymous codons that are more frequently used by that host.[7-11] This strategy has been used to improve protein yield by increasing the compatibility between the introduced gene and the host translation system. However, maximizing the use of host-preferred codons does not necessarily improve protein expression or protein quality. The aggressive codon optimization can substantially alter the native distribution of frequent and rare codons along the coding sequence, which may be undesirable because codon usage is not only related to overall translation efficiency, but can also influence local translation elongation rates. In native genes, for instance, the regions enriched in rare or less-preferred codons may help create the ribosomal pauses that contribute to co-translational protein folding, domain formation, or proper maturation of the nascent polypeptide.[12-14] Therefore, disrupting these native-like codon-usage patterns may negatively affect the structural and functional integrity of the encoded protein, even when the amino acid sequence remains unchanged.

To address the limitations, codon harmonization has emerged as an alternative strategy for designing genes for heterologous expression.[5,15-18] Unlike the conventional codon optimization, which generally aims to maximize the use of codons preferred by the host organism, codon harmonization seeks to preserve the relative pattern of codon usage found in the native gene. In codon harmonization, codons are selected so that the regions predicted to be translated relatively rapidly or slowly in the native organism are expected to show analogous relative translation behavior in the expression host. By maintaining the native-like patterns of local codon usage, codon harmonization aims to adapt a coding sequence to the host translation system while retaining the translation dynamics that may be important for proper protein folding, stability, and function.

To quantify the local codon-usage patterns, codon harmonization frameworks commonly rely on profile-based metrics.[19-21] One widely used metric is $\%MinMax$, which quantifies the extent to which the codons within a local sequence "window" are biased toward frequently or infrequently used synonymous codons in a reference organism (see **Section 3.2** for details).[21] To construct a $\%MinMax$ profile, a $\%MinMax$ value is computed for a sliding window at successive positions along the coding sequence, based on the usage frequencies of the codons within the window. Positive $\%MinMax$ values indicate the enrichment of relatively frequent synonymous codons, and vice versa. The position-dependent $\%MinMax$ profile, therefore, provides a representation of local codon usage along the sequence, capturing not only the overall codon composition of a gene but also the spatial arrangement of frequent and rare codons. $\%MinMax$ has been used both for the post-hoc evaluation of codon-usage patterns and as a profile-matching criterion in search-based codon harmonization algorithms.[17,18,21,22] In the algorithms, the design objective is to generate a heterologous coding sequence, the local codon-usage patterns of which resemble those of the native sequence, thereby preserving the native-like patterns of relative codon rarity and commonness.

In parallel, deep-learning-based approaches to codon design have recently attracted increasing attention. However, most existing neural or machine-learning-based methods have focused primarily on host adaptation, often by improving Codon Adaptation Index (CAI)-related metrics or by increasing the use of codons preferred by the expression host.[22-28] These approaches are useful for improving host compatibility, but they do not necessarily preserve the local pattern of codon usage that is central to codon harmonization, as aforementioned. In particular, the use of "differentiable" codon-harmonization objectives for gradient-based codon design remains relatively unexplored.

A major reason for this gap is computational. Conventional codon harmonization algorithms operate directly on discrete codon sequences. Because each position in the sequence is assigned a specific codon, $\%MinMax$ values can be calculated directly, and the resulting profile be compared with the profile of the native sequence. As a result, $\%MinMax$ is well suited for the search-based or other non-gradient optimization procedures. Neural codon design, however, typically represents the codon choices probabilistically during optimization. Instead of selecting one codon at each position from the beginning, the model may assign probabilities to multiple synonymous codons and update these probabilities through gradient-based learning. Therefore, the objective function must be differentiable with respect to codon probabilities or their underlying model parameters. The conventional $\%MinMax$ metric, denoted as $\%MinMax_{[h]}$ in this paper, is not naturally compatible with this setting, because $\%MinMax_{[h]}$ depends on the discrete codon identities and uses a hard branch between $\%Max$ and $\%Min$ normalization schemes. This limitation has motivated us to develop a differentiable codon-harmonization objective, which retains the interpretability of $\%MinMax$ while enabling neural sequence design. In this paper, we introduce $\boldsymbol{Smooth}$ $\boldsymbol{\%MinMax}$, denoted as $\boldsymbol{\%MinMax_{[s]}}$, as a differentiable relaxation of the conventional $\%MinMax_{[h]}$ metric for codon harmonization. The goal of $\%MinMax_{[s]}$ is to retain the biological interpretation of the original $\%MinMax_{[h]}$ profile while making it compatible with gradient-based neural sequence design.

$\%MinMax_{[s]}$ modifies two key components of the conventional metric. First, instead of assigning codon-usage values based on the discrete codon identities, $\%MinMax_{[s]}$ computes probability-weighted codon-usage values over the distribution of synonymous codons at each sequence position. This probability-weighted formulation allows the metric to evaluate codon-usage patterns even when codon choices are represented probabilistically during optimization. Second, $\%MinMax_{[s]}$ replaces the hard branch between $\%Max$ and $\%Min$ normalization with a smooth sigmoid-gated interpolation. This sigmoid-gated formulation preserves the signed interpretation of $\%MinMax_{[h]}$, while remaining differentiable. We show that $\%MinMax_{[s]}$ closely approximates $\%MinMax_{[h]}$ while providing the stable gradients suitable for profile-based codon harmonization. To evaluate this capability in a heterologous setting, we trained a model to design *E. coli* coding sequences for human proteins by conditioning on the amino acid sequence and target-profile context and minimizing a profile-matching loss. At inference, we employed profile-guided beam search, instead of the conventional greedy decoding.[29] Together, these results establish $\%MinMax_{[s]}$ as a differentiable bridge between classical profile-based codon harmonization and neural synonymous-sequence design.

# 2. Background & Related Work

## 2.1. Design Objectives and Evaluation Metrics in Codon Optimization

Codon optimization studies have used a wide range of objectives and evaluation metrics to guide synonymous-sequence design. Among these metrics, CAI is one of the most widely used measures of host adaptation. CAI was originally developed to quantify how closely a coding sequence followed the codon-usage characteristics of highly expressed genes in a reference organism.[30-33] Because highly expressed genes are often enriched for the codons that are efficiently translated by the host, CAI has become a common proxy for the compatibility between a designed coding sequence and the translational machinery of the expression host.

In addition to CAI, codon optimization frameworks often incorporate other sequence-level criteria, including GC content, codon-pair statistics, mRNA secondary-structure, avoidance of undesired sequence motifs, and additional constraints related to cloning, synthesis, or regulatory compatibility.[34-37] Together, these objectives reflect the fact that successful heterologous gene expression depends not only on codon preference, but also on broader properties of the coding sequence.

More recently, learning-based approaches have expanded the codon-design landscape by learning host-specific or natural-like codon preferences directly from sequence data. Examples include recurrent neural network (RNN)-based codon optimization models and Transformer-based multispecies codon design frameworks.[22-28] Some methods also consider rare-codon preservation or broader multi-objective optimization strategies that balance multiple sequence properties simultaneously.[38,39] These developments have increased the flexibility of codon design and enabled models to capture more complex sequence patterns than the traditional rule-based methods.

Despite these advances, most existing design objectives remain focused on improving host adaptation or satisfying global sequence-level constraints. They generally do not explicitly preserve the local arrangement of frequent and rare codons along the coding sequence. This distinction is important because local codon-usage patterns, rather than global codon composition alone, form the central design principle of codon harmonization.

### 2.2. Search-Based Codon Harmonization

Codon harmonization directly addresses this limitation by designing synonymous coding sequences that preserve native-like codon-usage patterns during heterologous expression. Instead of simply increasing the use of host-preferred codons, harmonization aims to reproduce the relative pattern of codon commonness and rarity observed in the original gene, but within the codon-usage context of the expression host.

Several search-based algorithms have been developed for this purpose. CHARMING formulates codon harmonization as the design of synonymous coding sequences that reproduce a target codon-usage profile in the expression host, providing a representative framework for profile-guided heterologous gene expression.[17] More recently, MOSAIC has introduced a Monte Carlo simulated-annealing approach for linked codon harmonization.[18] By extending the optimization space from individual codon substitutions to groups of linked codons, MOSAIC enables harmonization while accounting for local dependencies among neighboring codon choices.

## 3. Smooth $\%MinMax$

### 3.1. Problem Setup for Synonymous-Codon Sequence Design

Let $A = (a(1), \ldots, a(T))$ denote an amino-acid sequence of length $T$. For each amino acid $a(i)$ at position $i$, let $\mathcal{S}(a(i))$ denote the set of synonymous codons that encode $a(i)$. The problem of synonymous codon sequence design is to construct a coding sequence $\mathcal{C} = (c_1, \ldots, c_T)$ subject to $c_i \in \mathcal{S}(a(i))$.

Let $Y$ denote the target profile and $F(\mathcal{C})$ the profile calculated from coding sequence $\mathcal{C}$. The problem can be written as

$$\mathcal{C}^* = \underset{\mathcal{C}: c_i \in \mathcal{S}(a_i)}{\operatorname{argmin}} \mathcal{L}(F(\mathcal{C}), Y),$$

where $\mathcal{L}$ is a loss function. $\%MinMax_{[h]}$ defines $F(\mathcal{C})$ on discrete codon sequences, as described in **Section 3.2**. **Section 3.3** introduces a differentiable formulation, $\%MinMax_{[s]}$, which enables this profile-matching objective to be optimized with gradient-based methods.

### 3.2. Hard $\%\boldsymbol{MinMax}$

The $\%MinMax_{[h]}$ metric quantifies local deviations in codon usage relative to the average usage of synonymous codons. At each amino acid position in a given protein sequence, the usage frequency of the observed codon is compared with the average usage frequency of all synonymous codons encoding the same amino acid. The resulting deviations are then summed over a sliding window to produce a local codon-usage profile.

Let $\mathcal{S}(a)$ denote the set of synonymous codons encoding amino acid $a$, and let $u(c)$ denote the usage frequency of codon $c$ in the reference organism. The usage frequency is defined here as the number of occurrences of a given codon per 1,000 codons in the coding sequences of a given organism. For each amino acid $a$, the average synonymous-codon usage frequency is defined as

$$\mu_a = \frac{1}{|\mathcal{S}(a)|} \sum_{c \in \mathcal{S}(a)} u(c),$$

and the maximum and minimum synonymous-codon usage frequencies are defined as

$$u_a^{\max} = \max_{c \in \mathcal{S}(a)} u(c), \qquad u_a^{\min} = \min_{c \in \mathcal{S}(a)} u(c).$$

Let

$$c_i^{(k)} \in \mathcal{S}(a(i))$$

denote the selected synonymous codon at position $i$, where $k$ indexes the codon within the synonymous-codon set. The corresponding codon-usage value is defined as

$$x_i^{(k)} = u(c_i^{(k)}).$$

The residue-level deviation from the average synonymous-codon usage is then defined as

$$\delta_i = x_i^{(k)} - \mu_{a(i)}.$$

To obtain a local profile, these deviations are aggregated over a sliding window. Let $W_t$ denote the set of positions included in the sliding window starting at position $t$. The window-level deviation is defined as

$$\Delta_t = \sum_{i \in W_t} \delta_i.$$

The maximum possible positive deviation within window $W_t$ is

$$D_t^+ = \sum_{i \in W_t} (u_{a(i)}^{\max} - \mu_{a(i)}),$$

and the maximum possible negative deviation within window $W_t$ is

$$D_t^- = \sum_{i \in W_t} (\mu_{a(i)} - u_{a(i)}^{\min}).$$

In the $\%MinMax_{[h]}$ formulation, the normalization denominator is selected by a hard branch. If the window-level deviation is non-negative, $D_t^+$ is used; if the deviation is negative, $D_t^-$ is used. The $\%MinMax_{[h]}$ value at position $t$ is then computed as

$$M_t = \begin{cases} 100 \cdot \dfrac{\Delta_t}{D_t^+}, & \Delta_t \geq 0, \\ 100 \cdot \dfrac{\Delta_t}{D_t^-}, & \Delta_t < 0. \end{cases}$$

### 3.3. Smooth $\%MinMax$

Let $z_i^{(k)}$ denote the learnable pre-softmax score associated with codon $c_i^{(k)}$ at position $i$. We refer to these pre-softmax scores as logits in the machine-learning sense. These logits are optimized directly during gradient-based codon-sequence design, rather than treated as fixed outputs of a pretrained model.

The probability of selecting codon $c_i^{(k)}$ is defined by applying a softmax over $\mathcal{S}(a(i))$:

$$p_i\left(c_i^{(k)}\right) = \frac{\exp\left(z_i^{(k)}\right)}{\sum_{c_i^{(l)} \in \mathcal{S}(a(i))} \exp\left(z_i^{(l)}\right)}$$

Thus,

$$p_i\left(c_i^{(k)}\right) \geq 0, \qquad \sum_{c_i^{(k)} \in \mathcal{S}(a(i))} p_i\left(c_i^{(k)}\right) = 1.$$

The discrete codon-usage value is then replaced by a probability-weighted average of synonymous-codon usage values. Specifically, the relaxed codon-usage value at position $i$ is defined as

$$\tilde{x}_i = \sum_{c_i^{(k)} \in \mathcal{S}(a(i))} p_i\left(c_i^{(k)}\right) u(c_i^{(k)}).$$

This quantity generalizes the observed codon-usage value used in $\%MinMax_{[h]}$. When

$$p_i\left(c_i^{(k)}\right)$$

is one-hot over $\mathcal{S}(a(i))$, $\tilde{x}_i$ is exactly equal to the usage frequency of the selected codon. When the distribution is soft, $\tilde{x}_i$ is determined by the full probability distribution over $\mathcal{S}(a(i))$, rather than by a single selected codon, and remains differentiable with respect to the corresponding logits.

As in $\%MinMax_{[h]}$, this relaxed codon-usage value is compared with the average synonymous-codon usage frequency for the same amino acid. The residue-level relaxed deviation is defined as

$$\tilde{\delta}_i = \tilde{x}_i - \mu_{a(i)}.$$

A positive $\tilde{\delta}_i$ indicates that the probability mass over $\mathcal{S}(a(i))$ is biased toward synonymous codons with above-average usage frequencies. Conversely, a negative $\tilde{\delta}_i$ indicates that the probability mass is biased toward synonymous codons with below-average usage frequencies.

The window-level relaxed deviation is

$$\tilde{\Delta}_t = \sum_{i \in W_t} \tilde{\delta}_i.$$

The positive and negative normalization terms, $D_t^+$ and $D_t^-$, are defined as in $\%MinMax_{[h]}$, but $\%MinMax_{[s]}$ replaces the hard branch with a sigmoid gate:

$$g_t = \sigma(\beta\widetilde{\Delta}_t) = \frac{1}{1 + \exp(-\beta\widetilde{\Delta}_t)},$$

where the hyperparameter $\beta$ controls the sharpness of the transition. If $\widetilde{\Delta}_t$ is strongly positive, then $g_t$ approaches 1. If $\widetilde{\Delta}_t$ is strongly negative, then $g_t$ approaches 0. Around $\widetilde{\Delta}_t = 0$, the gate smoothly interpolates between the positive and negative branches. The smooth denominator is therefore defined as

$$\widetilde{D}_t = g_t D_t^+ + (1 - g_t) D_t^-.$$

Finally, the $\%MinMax_{[s]}$ value at position $t$ is computed as

$$\widetilde{M}_t = 100 \cdot \frac{\widetilde{\Delta}_t}{\widetilde{D}_t}.$$

This formulation preserves the signed interpretation of $\%MinMax_{[h]}$. Positive values indicate the local enrichment of codons with higher usage frequencies, whereas negative values indicate the local enrichment of codons with lower usage frequencies. The key difference is that all steps are differentiable with respect to the learnable pre-softmax scores through the synonymous-codon probabilities.

### 3.4. Effect of the Smoothing Parameter $\beta$

The smoothing parameter $\beta$ determines the sharpness of the sigmoid-gated transition between the positive and negative normalization branches in $\%MinMax_{[s]}$, thus controlling the degree to which the smooth formulation approximates the hard branch selection. As shown in Figure 1a, the sigmoid gate $\sigma(\beta\widetilde{\Delta}_t)$ changes from a shallow S-shaped curve to an increasingly sharp step as $\beta$ increases. Smaller values of $\beta$ produce a gradual transition around $\widetilde{\Delta}_t = 0$, allowing both $D_t^+$ and $D_t^-$ to contribute over a wider range of deviations, which increases smoothness but can introduce larger discrepancies from $\%MinMax_{[h]}$. Larger values of $\beta$ make the sigmoid gate increasingly step-like, thereby improving agreement with the hard formulation but concentrating the transition within a narrower interval around the branch boundary. Therefore, $\beta$ mediates a trade-off between smoothness and fidelity to $\%MinMax_{[h]}$.

We assessed this trade-off independently of neural codon optimization. The purpose of this analysis was to verify that $\%MinMax_{[s]}$ provided a faithful differentiable approximation of the conventional hard profile under controlled conditions. We used a human coding-sequence dataset, derived from the MoSAIC dataset after excluding the sequences shorter than 30 codons or longer than 2048 codons. For each coding sequence, we computed the conventional $\%MinMax_{[h]}$ profilefrom the discrete codon identities, with the widow size of 10, using the specified codon-usage table and sliding-window configuration. We then represented the same sequence as a one-hot synonymous-codon distribution at each amino-acid position and computed the corresponding $\%MinMax_{[s]}$ profile under identical settings. Because the one-hot representation makes the relaxed codon-usage value exactly equal to the usage frequency of the selected codon, any difference between $\%MinMax_{[s]}$ and $\%MinMax_{[h]}$ is attributable solely to the sigmoid-gated denominator.

Approximation accuracy was evaluated across different values of $\beta$ using mean absolute error (MAE), mean squared error (MSE), and Pearson correlation between the smooth and hard profiles. Increasing $\beta$ improved agreement with $\%MinMax_{[h]}$ across all three metrics. The mean MAE decreased from 4.096±1.368 at $\beta = 0.001$ to 0.0129±0.0088 at $\beta = 1$ (Figure 1b), and the mean MSE decreased from 36.598±22.950 to 0.00283±0.00382. The Pearson correlation increased from 0.981±0.0121 to 0.999998±0.000002, indicating near-identical profile shapes at large $\beta$.

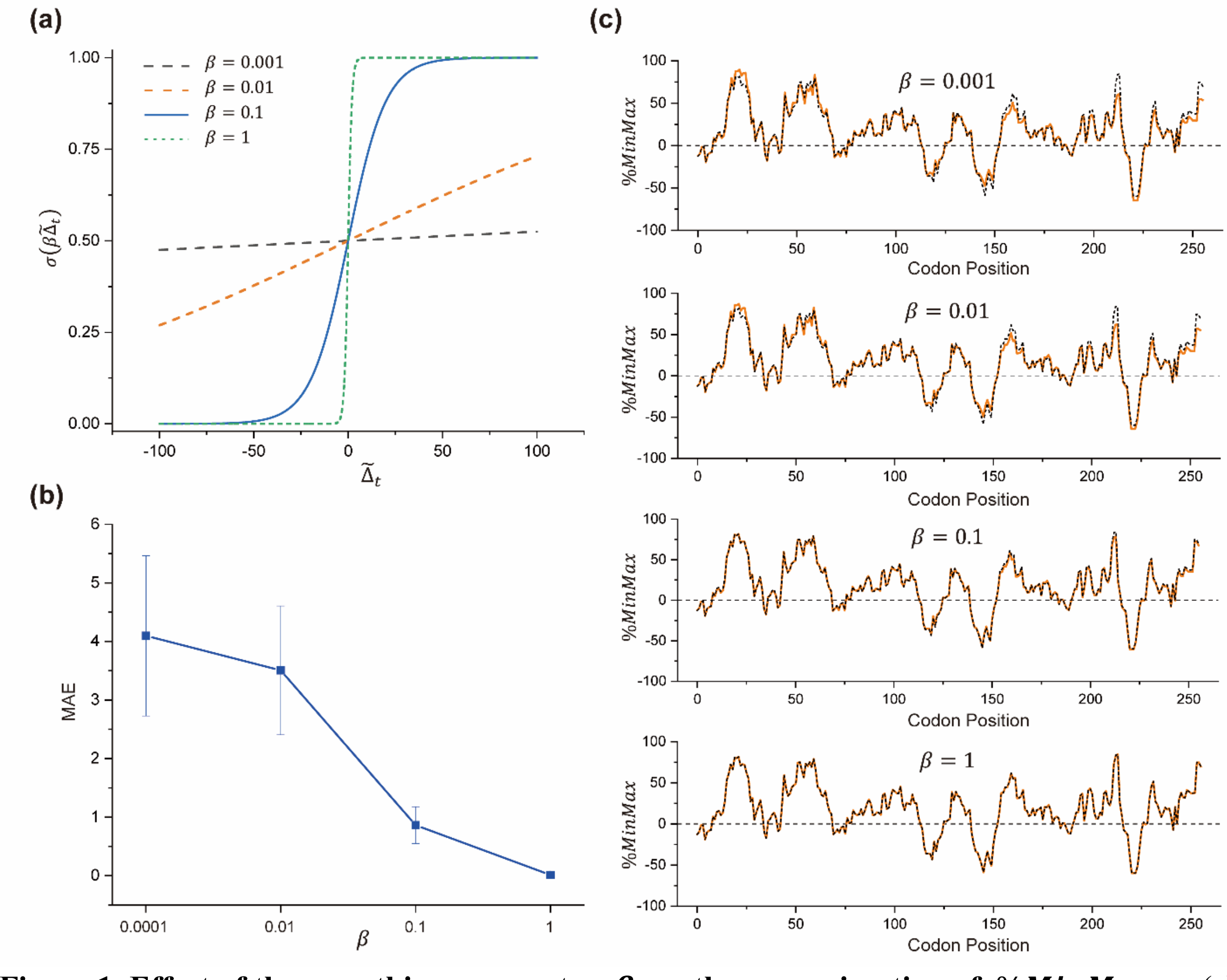

**Figure 1. Effect of the smoothing parameter $\beta$ on the approximation of $\%MinMax_{[h]}$.** (a) Sigmoid gating function, $\sigma(\beta\widetilde{\Delta}_t)$, evaluated over the range of $\widetilde{\Delta}_t$ for $\beta = 0.001, 0.01, 0.1,$ and 1. (b) MAE values between $\%MinMax_{[s]}$ and $\%MinMax_{[h]}$ profiles as a function of $\beta$. Error bars indicate the standard deviation across 20,070 human coding sequences. (c) Representative profile comparisons for the four values of $\beta$. The $\%MinMax_{[h]}$ profile is shown as a black dashed line, and the corresponding $\%MinMax_{[s]}$ profile is shown as an orange solid line.

Visual inspection of representative profiles supported these quantitative results (Figure 1c). Small $\beta$ values produced visible deviations from the hard profile, particularly near sharp peaks and locally fluctuating regions. These discrepancies decreased as $\beta$ increased. At $\beta = 0.1$, the smooth and hard profiles were nearly indistinguishable by visual inspection, while the sigmoid transition remained appreciably smoother than at $\beta = 1$. Although $\beta = 1$ gave the closest numerical approximation to $\%MinMax_{[h]}$, it also made the gate approximately step-like, reducing the practical benefit of smoothing for gradient-based optimization. We, therefore, selected $\beta = 0.1$ for subsequent experiments, as it provided a practical compromise between close approximation to the hard metric and preservation of a smooth transition around $\widetilde{\Delta}_t = 0$.

In this study, $\beta$ was treated as a fixed hyperparameter. However, the same formulation could be extended to a learnable smoothing parameter. To ensure positivity, $\beta$ could be parameterized as:

$$\beta = \mathrm{softplus}(\eta) + \beta_{\min},$$

where $\eta$ is an unconstrained learnable scalar, and $\beta_{\min} > 0$ is a small lower bound. This approach would allow the sharpness of the sigmoid gate to be optimized jointly with the synonymous-codon logits. We do not pursue the extension in this work.

# 4. Experiments

We evaluated whether $\%MinMax_{[s]}$ could serve as a differentiable objective for neural codon harmonization. For the proof-of-concept analysis, we used the human coding-sequence dataset previously employed in MoSAiC.[18] The dataset comprised 20,070 protein-coding sequences derived from T2T-CHM13, a reference genome assembly of *Homo sapiens*. Sequences shorter than 30 amino acids or longer than 2,048 amino acids were excluded. The remaining sequences were partitioned into chunks of up to 512 amino acids, with an overlap equal to the $\%MinMax$ window size of 10 between adjacent chunks. This overlap ensured that every valid $\%MinMax$ value, including those near chunk boundaries, could be computed without loss of sequence context. After preprocessing, the dataset contained 28,474 sequence chunks, which were randomly divided into training and test sets at a ratio of 9:1, yielding 25,626 training samples and 2,848 test samples, respectively.

As an initial implementation, we used a deliberately simple neural codon-design model to test whether $\%MinMax_{[s]}$ could support profile-guided optimization (Figure 2a). The input comprised an amino acid sequence and a target $\%MinMax_{[h]}$ profile computed from the corresponding human coding sequence. The amino acid sequence was first mapped to learnable embeddings and processed by two bidirectional long short-term memory (BiLSTM) modules, which generated a context-dependent representation at each sequence position. The resulting hidden representation was concatenated with a local segment of the target $\%MinMax_{[h]}$ profile and passed through a prediction head to produce the synonymous-codon logits for the amino acid at that position. The model then parameterized a probability distribution over the synonymous codons encoding that amino acid, with the corresponding logits serving as the learnable variables for gradient-based optimization. During optimization, these logits were converted into codon probabilities using a masked softmax function, ensuring that only synonymous codons for each amino acid could have nonzero probabilities. The $\%MinMax_{[s]}$ profile was then computed from these probabilities using *E. coli* codon-usage frequencies, so that the optimized sequence was evaluated in the codon-usage context of the intended expression host.

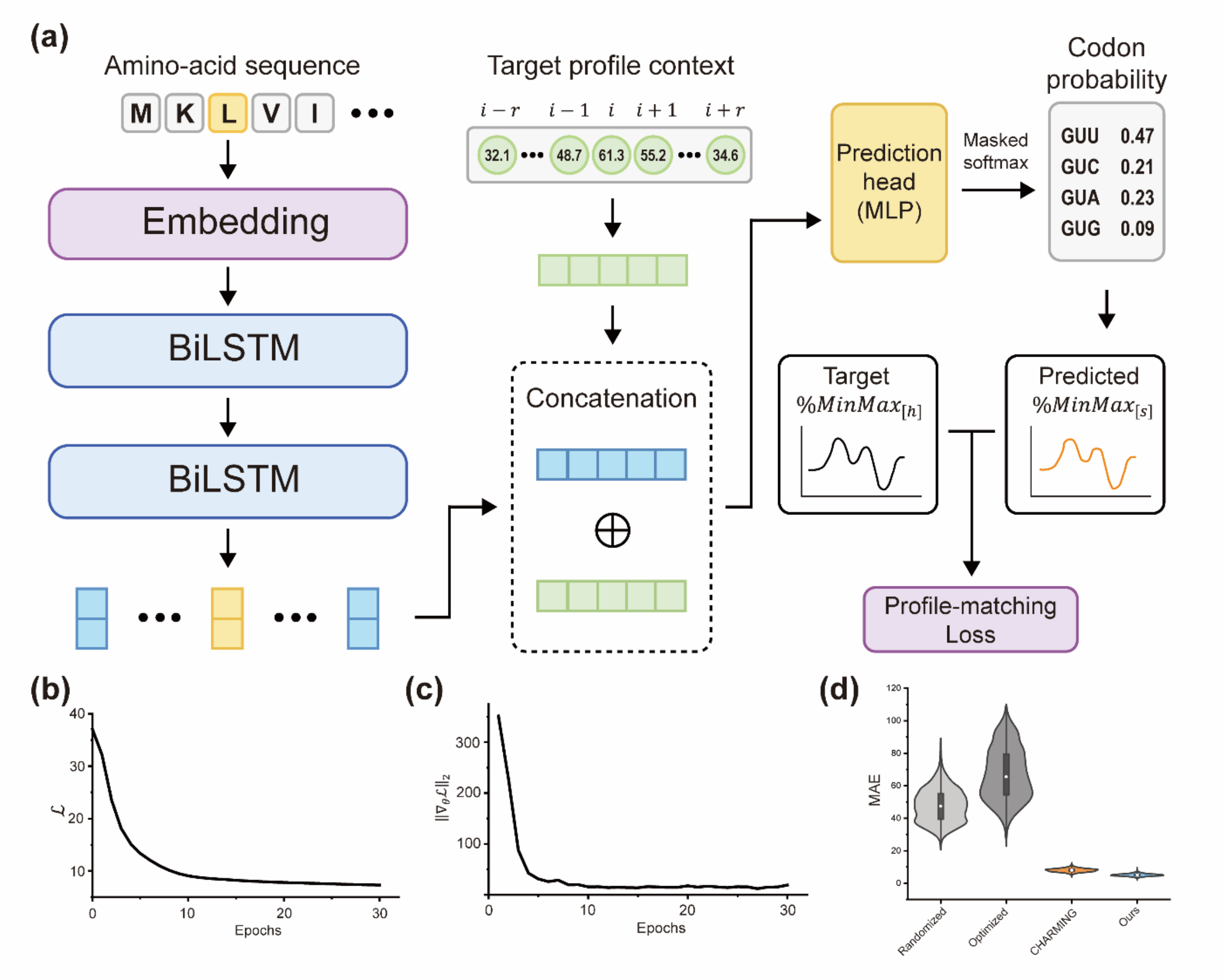

**Figure 2. Neural codon harmonization using $\%MinMax_{[s]}$ as a differentiable profile-matching objective.** (a) Overview of the model architecture and training procedure. (b) Profile-matching loss $\mathcal{L}$ over training epochs. MLP: multilayered perceptron. (c) Euclidean norm of the loss gradient with respect to the model parameters, $\|\nabla_\theta \mathcal{L}\|_2$, over training epochs. (d) Distribution of profile MAE across methods for human-to-*E. coli* heterologous codon harmonization. Lower values indicate closer agreement with the target profile. MAE: mean absolute error.

The optimization objective was to match the *E. coli* $\%MinMax_{[s]}$ profile to the human-derived target $\%MinMax_{[h]}$ profile by minimizing a profile-matching loss between the smooth *E. coli* profile and the target profile. The loss function was defined as the MAE between the two profiles:

$$\mathcal{L} = \sum_{t=0}^{T-|W_t|} \left|\widetilde{M}_t - M_t\right|,$$

where $\widetilde{M}_t$ is a $\%MinMax_{[s]}$ value, and $M_t$ is the target $\%MinMax_{[h]}$ value at the position $t$. During training, model parameters were updated iteratively by minimizing the mini-batch average of this sequence-level profile-matching loss. This setup was designed to test whether $\%MinMax_{[s]}$ could guide the gradient-based optimization toward the *E. coli* synonymous-codon distributions that preserve the local codon-usage pattern of the original human sequence.

Furthermore, the local target-profile segment was defined by a profile-context radius $r$. For a given position $i$, the target values from positions $i - r$ to $i + r$ were provided when predicting the codon at position $i$. The baseline model used only the target value at the corresponding position, equivalent to $r = 0$. We first performed preliminary optimization experiments using the profile-context radii from 1 to 5 to examine whether broader local context from the target profile improved profile matching. The maximum radius was set to 5 because, for the $\%MinMax_{[h]}$ window size of 10 codons used in this experiment, the codon selected at a given position influences profile values within an approximately five-position neighborhood. Based on the pre-screening results, we used $r = 5$ for the subsequent optimization analyses.

To verify that $\%MinMax_{[s]}$ functioned as a differentiable training objective, we monitored both the profile-matching loss and the norm of its gradient. As shown in Figure 2b, the training MAE decreased progressively from an initial value of 37.0376 to 7.3033, demonstrating that the gradient-based updates improved the alignment between the predicted smooth *E. coli* profile and the human-derived target profile. On the other hand, the gradient norm was defined as

$$\|\nabla_\theta \mathcal{L}\|_2 = \sqrt{\sum_i \left\|\frac{\partial \mathcal{L}}{\partial \theta_i}\right\|_2^2},$$

where $\theta$ denotes the model parameters. The gradient norm remained nonzero during training (Figure 2c), confirming that the profile-matching loss propagated gradient signals through the $\%MinMax_{[s]}$ computation to the model parameters. Additionally, the $\%MinMax_{[s]}$ profiles for the test set closely agreed with the corresponding $\%MinMax_{[h]}$ profiles (Figure S1).

We then evaluated whether the $\%MinMax_{[s]}$ could be applied to heterologous codon harmonization of human coding sequences for expression in *E. coli*. The model received human protein sequences and their target $\%MinMax_{[h]}$ profiles as input, and was trained using a $\%MinMax_{[s]}$-based profile-matching loss. At inference, profile-guided beam search was used to generate synonymous coding sequences that closely matched the target profiles. We compared our method with three baselines: random synonymous-codon selection, frequency-based codon optimization, and CHARMING. The randomized baseline selected a synonymous codon randomly at each position, whereas the frequency-based optimization baseline selected the most frequent *E. coli* codon for each amino acid. For each method, we computed the MAE between the target profile and the $\%MinMax_{[h]}$ profiles of each generated sequence. The MAE distribution for our method was concentrated at lower values than those of the baselines, demonstrating its superior ability for accurately preserving the target profiles (Figure 2d).

Together, these results indicated that $\%MinMax_{[s]}$ largely preserved the signed profile structure of the conventional $\%MinMax_{[h]}$ metric while remaining usable as a gradient-based optimization objective. The reduction in training loss showed that the $\%MinMax_{[s]}$-based profile-matching objective could be minimized by gradient-based optimization, and the test-set profile agreement supported its applicability beyond the training sequences. After optimization, the resulting *E. coli* synonymous-codon probability distributions produced the smooth profiles that more closely matched the human-derived target profiles. More importantly, when converted into discrete synonymous coding sequences by profile-guided beam search, these optimized distributions yielded $\%MinMax_{[h]}$ profiles that more closely matched the human-derived target profiles than those generated by the baseline methods. These findings support the feasibility of using $\%MinMax$ as a differentiable objective for probabilistic neural codon harmonization.

# 5. Conclusions

In this study, we introduced $Smooth\ \%MinMax$, denoted $\%MinMax_{[s]}$, as a differentiable relaxation of the conventional hard $\%MinMax$ metric, denoted $\%MinMax_{[h]}$, for profile-based neural codon harmonization. By replacing discrete codon-usage values with probability-weighted synonymous-codon usage values and replacing the hard $\%Max/\%Min$ branch with a sigmoid-gated interpolation, $\%MinMax_{[s]}$ retains the signed interpretation of the original metric while making the codon-usage profile differentiable with respect to synonymous-codon probabilities and learnable logits. This formulation provides a way to incorporate codon-harmonization profiles directly into gradient-based neural codon design, addressing a key incompatibility between classical profile-based harmonization metrics and probabilistic sequence-generation models.

As a proof-of-concept, we applied $\%MinMax_{[s]}$ to human-to-*E. coli* neural codon harmonization and showed that the smooth metric could approximate $\%MinMax_{[h]}$, support the gradient-based profile matching, and guide optimized synonymous-codon probability distributions toward human-derived target profiles in the *E. coli* codon-usage context. Profile-guided beam search further bridged the continuous-to-discrete gap by converting the learned codon-probability distributions into discrete coding sequences while preserving the target profiles. These findings support the feasibility of $\%MinMax_{[s]}$ as a differentiable objective for neural codon harmonization. However, this study was limited to an initial probabilistic design framework and did not incorporate additional biological constraints or experimentally evaluate the resulting designs. Nonetheless, $\%MinMax_{[s]}$ provides a practical foundation for integrating conventional codon harmonization with neural sequence design.

## 6. Code Availability

The source code and data used in this study are available at: https://github.com/CIS-group/SmoothMinMax.

# SUPPORTING INFORMATION

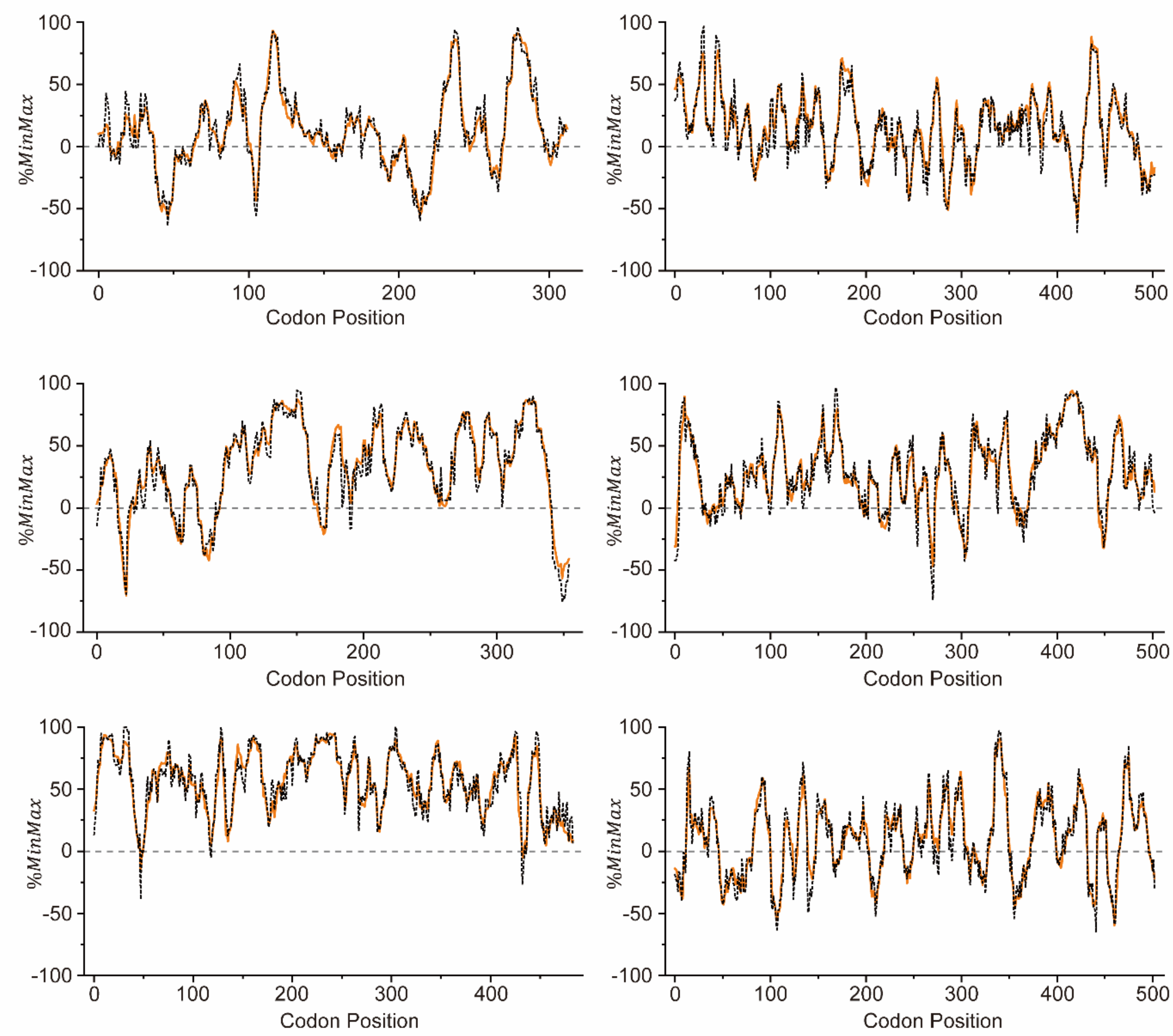

**Figure S1. Six representative profiles randomly sampled from the test set.** The target $\%MinMax_{[h]}$ profile is shown as a black dashed line, and the predicted $\%MinMax_{[s]}$ profile is shown as an orange solid line.